\documentstyle[12pt]{article}

\newcommand{\eqn}[1]{(\ref{#1})}

\newcommand{\complex}{{\bb C}} 
\newcommand{\complexs}{{\bbs C}} 
\newcommand{\zed}{{\bb Z}} 
\newcommand{\real}{{\bb R}} 
\newcommand{\zeds}{{\bbs Z}} 
\newcommand{\id}{{\bb I}} 
\newcommand{\alg}{{\cal A}} 

\font\mybb=msbm10 at 12pt
\def\bb#1{\hbox{\mybb#1}}
\font\mybbs=msbm10 at 9pt
\def\bbs#1{\hbox{\mybbs#1}}

\def\e{{\rm e}}
\def\semiprod{{\supset\!\!\!\!\!\!\!\times~}}

\def\beq{\begin{equation}}
\def\eeq{\end{equation}}
\def\bea{\begin{eqnarray}}
\def\eea{\end{eqnarray}}
\def\bd{\begin{displaymath}}
\def\ed{\end{displaymath}}

\makeatletter
\newdimen\normalarrayskip              
\newdimen\minarrayskip                 
\normalarrayskip\baselineskip
\minarrayskip\jot
\newif\ifold             \oldtrue            \def\new{\oldfalse}
\def\arraymode{\ifold\relax\else\displaystyle\fi} 
\def\@arrayskip{\ifold\baselineskip\z@\lineskip\z@
     \else
     \baselineskip\minarrayskip\lineskip2\minarrayskip\fi}
\def\@arrayclassz{\ifcase \@lastchclass \@acolampacol \or
\@ampacol \or \or \or \@addamp \or
   \@acolampacol \or \@firstampfalse \@acol \fi
\edef\@preamble{\@preamble
  \ifcase \@chnum
     \hfil$\relax\arraymode\@sharp$\hfil
     \or $\relax\arraymode\@sharp$\hfil
     \or \hfil$\relax\arraymode\@sharp$\fi}}
\def\@array[#1]#2{\setbox\@arstrutbox=\hbox{\vrule
     height\arraystretch \ht\strutbox
     depth\arraystretch \dp\strutbox
     width\z@}\@mkpream{#2}\edef\@preamble{\halign \noexpand\@halignto
\bgroup \tabskip\z@ \@arstrut \@preamble \tabskip\z@ \cr}%
\let\@startpbox\@@startpbox \let\@endpbox\@@endpbox
  \if #1t\vtop \else \if#1b\vbox \else \vcenter \fi\fi
  \bgroup \let\par\relax
  \let\@sharp##\let\protect\relax
  \@arrayskip\@preamble}
\makeatother

\newcommand{\newsection}[1]
{\vspace{4mm}
\pagebreak[3]
\addtocounter{section}{1}
\setcounter{equation}{0}
\setcounter{subsection}{0}
\begin{flushleft}
{\large\bf \thesection. #1}
\end{flushleft}
\nopagebreak
\medskip
\nopagebreak}

\newlength{\extraspace}
\newlength{\extraspaces}
\begin{document}

\renewcommand{\footnotesize}{\small}

\addtolength{\baselineskip}{.7mm}

\thispagestyle{empty}

\begin{flushright}
\baselineskip=12pt
DSF/51-97\\
OUTP-97-75P\\
hep-th/9712206\\
\hfill{  }\\
\hfill{  }\\ December 1997
\end{flushright}

\begin{center}

\baselineskip=12pt

{\Large\bf{Noncommutative Geometry and \\Spacetime Gauge Symmetries of
String Theory\footnote{Invited paper to appear in the special issue of the
Journal of {\it Chaos, Solitons and Fractals} on ``Superstrings, M, F, S,
$\dots$ Theory" (M.S. El Naschie and C. Castro, editors).}}}\\[8mm]

{\sc Fedele Lizzi}\footnote{E-mail: {\tt lizzi@na.infn.it}}
\\[4mm]
{\it Dipartimento di Scienze Fisiche, Universit\`a di Napoli Federico II\\ and
INFN, Sezione di Napoli, Mostra d'Oltremare Pad.~19, 80125 Napoli, Italy}
\\[6mm]
{\sc Richard J.\ Szabo\footnote{E-mail: {\tt r.szabo1@physics.oxford.ac.uk}}}
\\[4mm]
{\it Department of Physics -- Theoretical Physics, University of Oxford\\ 1
Keble Road, Oxford OX1 3NP, U.K.}
\\[40mm]

{\bf Abstract}

\begin{center}
\begin{minipage}{15cm}

We illustrate the various ways in which the algebraic framework of
noncommutative geometry naturally captures the short-distance spacetime
properties of string theory. We describe the noncommutative spacetime
constructed from a vertex operator algebra and show that its algebraic
properties bear a striking resemblence to some structures appearing in M
Theory, such as the noncommutative torus. We classify the inner automorphisms
of the space and show how they naturally imply the conventional duality
symmetries of the quantum geometry of spacetime. We examine the problem of
constructing a universal gauge group which overlies all of the dynamical
symmetries of the string spacetime. We also describe some aspects of toroidal
compactifications with a light-like coordinate and show how certain generalized
Kac-Moody symmetries, such as the Monster sporadic group, arise as gauge
symmetries of the resulting spacetime and of superstring theories.

\end{minipage}
\end{center}

\end{center}

\vfill
\newpage
\pagestyle{plain}
\setcounter{footnote}{0}
\setcounter{page}{1}
\stepcounter{subsection}

\newsection{Symmetries in Noncommutative Geometry}

The description of spacetime and its symmetries at very short length scales is
one of the main challenges facing theoretical physics at the end of this
millenium. At distances much larger than the Planck length, spacetime is
described by a differentiable manifold, whose symmetries are diffeormophisms
which describe the gravitational interactions. Additional interactions are
described by gauge theories, associated with the symmetries of some internal
structure living in the spacetime. The most promising candidate for a
description of spacetime and its symmetries at the quantum level is string
theory. Here the Planck length appears as the square root of the string
tension, and the coordinates of spacetime are target space fields, defined on a
two-dimensional world-sheet, which describe the surface swept out by the string
in spacetime.

In the case of a toroidally
compactified target space $T^d\cong\real^d/2\pi\Gamma$, with $\Gamma$ a
Euclidean rational lattice of rank $d$ with inner product $g_{\mu\nu}$, the
string is described by the Fubini-Veneziano fields\footnote{In this paper we
will consider only the simplest non-trivial case of closed strings with
world-sheet which is an infinite cylinder with local coordinates
$(\tau,\sigma)\in\real\times S^1$, and we set $z_\pm=\e^{-i(\tau\pm\sigma)}$.}
\beq
X^\mu_\pm(z_\pm)=x^\mu_\pm+ig^{\mu\nu}p_\nu^\pm\log z_\pm
+\sum_{k\neq0}\frac1{ik}~\alpha^{(\pm)\mu}_k~z_\pm^{-k}
\label{chiralmultfields} \ ,\label{FubVen}\eeq
where the zero-modes $x^\mu_\pm$ (the center of mass coordinates of the string)
and the (center of mass) momenta $p_\mu^\pm=\frac1{\sqrt2}(p_\mu\pm
d_{\mu\nu}^\pm w^\nu)$ are canonically conjugate variables. Here
$\{p_\mu\}\in\Gamma^*$ (the lattice dual to $\Gamma$)
are the spacetime momenta and $\{w^\mu\}\in\Gamma$
are winding numbers representing the number of times that the string wraps
around the cycles of the torus. The background matrices
$d_{\mu\nu}^\pm=g_{\mu\nu}\pm\beta_{\mu\nu}$ are constructed from the spacetime
metric $g_{\mu\nu}$ and an antisymmetric instanton tensor $\beta_{\mu\nu}$. The
set of momenta $\{(p_\mu^+,p_\mu^-)\}$ along with the quadratic form
\beq
\langle p,q\rangle_\Lambda\equiv p_\mu^+g^{\mu\nu}q_\nu^+-p^-_\mu
g^{\mu\nu}q_\nu^-=p_\mu v^\mu+q_\mu w^\mu \ ,
\label{quadform}\eeq
where $q_\mu^\pm=\frac1{\sqrt2}(q_\mu\pm d_{\mu\nu}^\pm v^\nu)$, form an even
self-dual Lorentzian lattice $\Lambda=\Gamma^*\oplus\Gamma$ of rank $2d$ and
signature $(d,d)$ called the Narain lattice \cite{narain}. The
functions (\ref{chiralmultfields}) define chiral multi-valued quantum fields.
The oscillatory modes $\alpha_k^{(\pm)\mu}=(\alpha_{-k}^{(\pm)\mu})^*$ in
(\ref{chiralmultfields}) yield bosonic creation and annihilation operators
(acting on some vacuum states $|0\rangle_\pm$) which generate the Heisenberg
algebra
\beq
\left[\alpha^{(\pm)\mu}_k,\alpha^{(\pm)\nu}_m\right]=k~g^{\mu\nu}~
\delta_{k+m,0}\ .
\label{creannalg}\eeq

The fundamental continuous symmetries of these fields are target space
reparametrization invariance and conformal invariance of the world-sheet. In
addition there are a number of quantum symmetries, such as dualities, mirror
symmetries, world-sheet parity, etc. which we shall discuss in section 4.
The interactions of the strings are described by a vertex
operator algebra, which we study in the next section. These interactions are
crucial for the theory at short distances, where a description based on a
cylindrical world-sheet breaks down. We want, however, to recover the usual
classical spacetime in some low energy limit.

Although string theory is a dramatic generalization of quantum field theory,
its tools still lie within the realm of traditional differential geometry,
which utilizes sets of points, differentiable manifolds, and commuting or
anticommuting fields. Recently, however, a much more radical approach to
geometry has been pioneered by Connes and others which goes under the name of
Noncommutative Geometry \cite{Book}. The central idea of Noncommutative
Geometry is that since all separable topological spaces (for example
manifolds) are completely characterized by the commutative $C^*$-algebra of
continuous complex-valued functions defined on it, it may be useful to regard
this algebra as the algebra generated by the coordinates of the space.
Conversely, given a {\em commutative} $C^*$-algebra, it is possible to
construct, with purely algebraic methods, a topological space and recover,
again with purely algebraic methods, all of the topological information about
it.  One can therefore substitute the study of the geometry and topology of a
space in terms of the relations among sets of points with a purely algebraic
description in terms of $C^*$-algebras. The terminology {\em noncommutative}
geometry refers to the possibility of generalizing these concepts to the case
where the algebra is noncommutative. This would be the case of some sort of
space in which the coordinates do not commute, such as that predicted by the
recent matrix model realizations of M Theory \cite{witten,bfss}, but this is
only one possibility (and as we will see a rather reductive one).

A noncommutative manifold is therefore a space described by a noncommutative
$C^*$-algebra. Within the algebraic framework of noncommutative geometry, the
symmetries of spacetime have a natural interpretation at the algebraic level
\cite{connesauto}. The algebra which describes the manifold (or its
generalization) is unchanged as a whole under automorphisms, i.e. isomorphisms
of the algebra into itself, as they just reshuffle the elements without
changing the basic algebraic relations. In fact, for a manifold $M$, the group
of diffeomorphisms of the space is naturally isomorphic to the group of
automorphisms of the commutative algebra $\alg=C^\infty(M)$ of smooth functions
defined on it,
\beq
f(\phi(x))=\sigma_\phi(f)(x) \ ,
\eeq
where $\phi(x)$ is a diffeomorphism of $M$ and $\sigma_\phi(f)\equiv
f\circ\phi$ is an (outer) automorphism of $\alg$.

When the algebra $\alg$ is noncommutative, a new class of automorphisms
appears, the inner automorphims, which act as conjugation of the elements of
the algebra by a unitary element $u\in\alg$,
\beq
\sigma_u(f)=ufu^{-1} \ .
\eeq
They represent internal fluctuations of the noncommutative geometry
corresponding to rotations of the elements of $\alg$. If, for example, we
consider the algebra of smooth functions from a manifold $M$ into the space
$M(N,\complex)$ of $N\times N$ complex-valued matrices, then the group of inner
automorphisms coincides with the group of local gauge transformations of a
$U(N)$ gauge theory on $M$. In general, as is customary in theories with
noncommutative fields, the algebra of observables is the tensor product of the
infinite-dimensional algebra of functions on $M$ with a finite-dimensional
algebra (typically a matrix algebra). In noncommutative
geometry this algebra describes a manifold, represented by the
infinite-dimensional algebra, in which the points have some sort of internal
structure, described by the finite-dimensional algebra. This is the case
in the example of the standard model \cite{standard}, where the internal
structure is described by an algebra whose unimodular group is the gauge group
$SU(3)\times SU(2)\times U(1)$. The application of noncommutative geometry
to the standard model has had some successes, but probably its most promising
potential lies in considering more complicated structures than just the tensor
product of a commutative algebra with a matrix algebra.

We see therefore that in noncommutative geometry diffeomorphisms and gauge
transformations appear on a very similar footing. Both represent
transformations of the algebra which leave it invariant, the difference being
that the latter ones are automorphisms which are internally generated by
elements of the algebra and appear as transformations of the internal space. In
the following we will apply this formalism to the spacetime described by string
theory. We will show how the geometry and topology relevant for general
relativity must be embedded in a larger noncommutative structure in string
theory.

\newsection{Vertex Operator Algebras}

At very high energies, strings appear as genuine extended objects, so that at
small distance scales it may not be possible to localize the coordinates of
spacetime with definite certainty. A long-standing idea in string theory is
that, below the minimum distance determined by the finite size of the string
(usually taken to be the Planck length), the classical concepts of spacetime
geometry break down. It is therefore natural to assume that the spacetime
implied by string theory is described by some noncommutative algebra. An
explicit realization of this idea arises in D-brane field theory \cite{witten}
and M Theory \cite{bfss} where the noncommutative coordinates of spacetime are
encoded in a finite-dimensional matrix algebra as described in the previous
section. However, there is another approach which in essence describes the
``space" of interacting strings by employing the vertex operator algebra of the
underlying two-dimensional conformal quantum field theory
\cite{fg}--\cite{lslong}. Vertex operators describe the interactions of strings
and act on the string Hilbert space as insertions on the world-sheet
corresponding to the emission or absorption of string states. The idea behind
this approach is that, to a large extent, the stringy spacetime is determined
by its quantum symmetries.

The complicated algebraic properties of a vertex operator algebra lead to a
very complex noncommutative spacetime. The underlying conformal field theory
admits two mutually commuting chiral algebras $\alg^\pm$ of observables, namely
the operator product algebras of (anti-)holomorphic fields. Each chiral algebra
contains a representation ${\rm Vir}^\pm$ of the infinite-dimensional Virasoro
algebra whose generators $L_k^\pm$ satisfy the commutation relations
\beq
\left[L_k^\pm,L_m^\pm\right]=(k-m)L_{k+m}^\pm+\mbox{$\frac
c{12}$}\left(k^3-k\right)\delta_{k+m,0} \ ,
\label{vircomm}\eeq
where $c$ is the conformal anomaly. The corresponding symmetry group represents
the basic conformal invariance of the string theory. For the case of toroidally
compactified bosonic strings, the vertex operator algebra
$\alg=\alg^+\otimes_{\complexs\{\Lambda\}}\alg^-$ is constructed using the
operator-state correspondence of local quantum field theory (here
$\complex\{\Lambda\}$ denotes a twisting factor from the group algebra of the
double cover of $\Lambda\otimes_\zeds\complex$ \cite{lslong}). The Hilbert
space on which the quantum fields \eqn{chiralmultfields} act is
\beq
{\cal H}=L^2(T^d)^\Gamma\otimes{\cal F}^+\otimes{\cal F}^- \ ,
\label{hilbert}\eeq
where $L^2(T^d)^\Gamma=\bigoplus_{w^\mu\in\Gamma}L^2(T^d)$ is the $L^2$ space
associated with the center of mass modes of the strings, whose connected
components are labelled by the winding numbers and whose basis is denoted
$|p^+;p^-\rangle=\e^{-ip_\mu x^\mu}$ in each component. ${\cal F}^\pm$ are Fock
spaces generated by the oscillatory modes $\alpha_k^{(\pm)\mu}$, and the unique
vacuum state of $\cal H$ is $|{\rm
vac}\rangle\equiv|0;0\rangle\otimes|0\rangle_+\otimes|0\rangle_-$. To a typical
homogeneous state
\beq
|\psi\rangle=|q^+;q^-\rangle\otimes\mbox{$\prod_j$}\,r_\mu^{(j)+}
\alpha_{-n_j}^{(+)\mu}|0\rangle_+\otimes\mbox{$\prod_k$}\,r_\nu^{(k)-}
\alpha_{-m_k}^{(-)\nu}|0\rangle_+
\label{homstate}\eeq
of $\cal H$, there corresponds the vertex operator
\beq
V(\psi;z_+,z_-)=~:~i\,V_{q^+q^-}(z_+,z_-)~\mbox{$\prod_j\frac{r_\mu^{(j)+}}
{(n_j-1)!}\,\partial_{z_+}^{n_j}X_+^\mu~\prod_k\frac{r_\nu^{(k)-}}
{(m_k-1)!}\,\partial_{z_-}^{m_k}X_-^\nu$}~:
\label{highspin}\eeq
where $(q^+,q^-),(r^+,r^-)\in\Lambda$, $\psi\in{\cal H}^*$ denotes the operator
with
$|\psi\rangle=\psi|{\rm vac}\rangle$, and
\beq
V_{q^+q^-}(z_+,z_-)\equiv
V(\e^{-iq_\mu^+x_+^\mu-iq_\mu^-x_-^\mu}\otimes\id;z_+,z_-)=(-1)^{q_\mu
w^\mu}:~\e^{-iq_\mu^+X_+^\mu(z_+)-iq_\mu^-X_-^\mu(z_-)}~:
\label{tachyon}\eeq
are the fundamental tachyon vertex operators. Then the smeared operators
$V(\psi;f)=\int d^2z~V(\psi;z_+,z_-)f(z_+,z_-)$, with $f$ a Schwartz space test
function, are well-defined and densely-defined on $\cal H$ with
$|\psi\rangle=V(\psi;f)|{\rm vac}\rangle$. They realize explicitly the
non-locality property required of noncommuting spacetime coordinate fields.

The Hilbert space $\cal H$ can be graded by the conformal dimensions
$\Delta_q^\pm$ which are highest weights of the irreducible representations of
the Virasoro algebras ${\rm Vir}^\pm$. The Virasoro operators in the present
case are $L_k^\pm=\frac12\sum_{m\in\zeds}g_{\mu\nu}:\alpha_m^{(\pm)\mu}
\alpha_{k-m}^{(\pm)\nu}:$, with $\alpha_0^{(\pm)\mu}\equiv
g^{\mu\nu}p_\nu^\pm$. They generate a representation of the Virasoro
algebra \eqn{vircomm} of central charge $c=d$. The grading is defined on the
subspaces ${\cal H}_{\Delta_q}\subset\cal H$ of states \eqn{homstate} which are
highest weight vectors,
\beq
L_0^\pm|\psi\rangle=\Delta_q^\pm|\psi\rangle~~~~~~,~~~~~~
L_k^\pm|\psi\rangle=0~~~\forall k>0 \ ,
\label{highestwt}\eeq
where $\Delta^+_q=\frac12g^{\mu\nu}q_\mu^+q_\nu^++\sum_jn_j$ and
$\Delta_q^-=\frac12g^{\mu\nu}q_\mu^-q_\nu^-+\sum_km_k$. The corresponding
operator-valued distributions \eqn{highspin} are called primary fields.

The operators \eqn{tachyon} generate the tachyon states
$|q^+;q^-\rangle\otimes|0\rangle_+\otimes|0\rangle_-\in L^2(T^d)$. The string
oscillations vanish on this subspace of $\cal H$, and when $w^\mu=0$
(equivalently $q^+=q^-$) the operators \eqn{tachyon} projected onto this
subspace coincide with the basis elements $\e^{-iq_\mu x^\mu}$ of the
commutative algebra $C^\infty(T^d)$.\footnote{See \cite{lslett,lslong} for a
precise definition of this projection.} Thus the tachyon vertex operators
represent the subspace of the full noncommutative string spacetime which
corresponds to ordinary classical spacetime. This is the noncommutative
geometry version of the appearence of classical general relativity in a
low-energy limit (i.e. at distances much larger than the Planck length). In
string theory a low-energy regime is one in which the vibrational modes of the
string are negligible and no string modes wind around the spacetime. In this
limit strings become effectively point particles which are well-described by
ordinary quantum field theory.

The tachyon operators \eqn{tachyon} can be thought of as sorts of ``plane wave"
basis fields on the space of conformal field configurations. The tachyon states
are highest weight states of the level 2 $u(1)_+^d\oplus u(1)_-^d$ current
algebra \eqn{creannalg}, so that the entire Hilbert space can be built up from
the actions of the $\alpha_k$'s for $k<0$ on these states. This current algebra
represents the target space reparametrization symmetry of the string theory.
Thus the tachyon vertex operators in this sense span the noncommutative vertex
operator algebra $\alg$ of the string spacetime. Another important class of
vertex operators for us will be the graviton operators
\beq
V_{q^+q^-}^{\mu\nu}(z_+,z_-)=~:~i\,V_{q^+q^-}(z_+,z_-)\,\partial_{z_+}X_+^\mu
\,\partial_{z_-}X_-^\nu~:\ .
\label{graviton}\eeq
These operators represent the Fourier modes of the background matrices
$d_{\mu\nu}^\pm$ of the toroidal target space, and they create the graviton
state $|q^+;q^-\rangle\otimes\alpha_{-1}^{(+)\mu}|0\rangle_+\otimes
\alpha_{-1}^{(-)\nu}|0\rangle_-$ of polarization $(\mu\nu)$. The graviton
states represent the lowest-lying states with non-trivial string oscillations,
and as such they generate the smallest stringy excitations of the
commutative spacetime determined by the tachyon sector of $\alg$.

The algebraic relations of $\alg$ which characterize the noncommutativity of
the chiral algebras can be combined into a single relation known as the Jacobi
identity of the vertex operator algebra \cite{lslong,flm}. It can be regarded
as a combination of the classical Jacobi identity for Lie algebras and the
Cauchy residue formula for meromorphic functions. The operator product algebra
can be encoded through the algebraic relations among the tachyon operators
\cite{lslong}, for which we find the clock algebra
\beq
V_{q^+q^-}(z_+,z_-)V_{r^+r^-}(w_+,w_-)=\e^{-\pi i\,\langle
q,r\rangle_\Lambda}~V_{r^+r^-}(w_+,w_-)V_{q^+q^-}(z_+,z_-)
\label{clockalg}\eeq
for $\pm\,{\rm arg}~z_\pm>\pm\,{\rm arg}~w_\pm$. The algebra \eqn{clockalg} is
a generalization of one of the original examples of a noncommutative geometry,
the noncommutative torus \cite{Book,fgr,Rieffel}. For $d=2$ this algebra
describes the quotient of the 2-torus $T^2$ by the orbit of a free particle
whose velocity vector forms an angle $\theta$ with respect to the cycles of
$T^2$. When $\theta$ is irrational the motion is ergodic and dense in the
torus, and the resulting quotient is not a topological space in the usual
sense. An equivalent way to visualize this is to consider the rotations of a
circle. It is then possible to describe the space by the algebra of functions
on a circle together with the action of these rotations. Such an algebra
$\alg_\theta$ is generated by two elements $U$ and $V$ which obey
\beq
UV=\e^{-\pi i\theta}~VU \ .
\label{nctorus}\eeq
This algebra has also appeared in the recent matrix model descriptions of M
Theory \cite{bfss,fgr} and in the noncommutative geometry of four-dimensional
gauge theories \cite{lsem}. When $\theta=2K/N$, with $K$ and $N$ relatively
prime positive integers, the generators $U$ and $V$ can be represented by
$N\times N$ matrices. Then the quotient $\alg_\theta/{\cal I}_\theta$ by the
ideal ${\cal I}_\theta$ of $\alg_\theta$ generated by its center is isomorphic,
as a $C^*$-algebra, to the full matrix algebra $M(N,\complex)$. When $\theta$
is irrational the algebra $\alg_\theta$ is infinite-dimensional.

The clock algebra \eqn{clockalg} (or rather its smeared version) thus resembles
the algebra $\alg_{\{\theta_{q,r}\}}$ of the noncommutative $d$-torus. However,
because the vertex operator algebra $\alg$ is determined from the quantum field
algebra of the conformal field theory, it really represents some large,
infinite-dimensional generalization of this structure. It can be regarded as an
augmentation, in an appropriate limiting procedure $\alg\sim
M(\infty,\complex)$, of the matrix algebras which describe the symmetry group
of the 11-dimensional supermembrane and the low-energy dynamics of M Theory
determined by the large-$N$ limit of a supersymmetric matrix model \cite{bfss}.
The matrix model for M Theory can be viewed as a particular truncation of a set
of vertex operators to finite-dimensional $N\times N$ matrices such that the
large-$N$ limit recovers aspects of the full non-perturbative dynamics. For
$d=2$, this limiting regime should be taken as a double scaling limit
$K,N\to\infty$ with $\theta=2K/N$ approaching an irrational number. In M Theory
there are several conjectures about what the appropriate large-$N$ limit should
be \cite{bfss,bss}, and from the above point of view such limiting regimes
should result in a vertex operator algebra structure. It is in this way that
the vertex operator algebra represents the noncommutative coordinates of string
theory. Furthermore, this construction demonstrates the unity between gauge
theories and string theories, an important ingredient of the unified framework
of M Theory. It is also possible to prove that
$\alg_\theta\cong\alg_{1/\theta}\cong\alg_{\theta+1}$ \cite{Rieffel}. When
translated into the language of the vertex operator algebra, this implies
certain duality symmetries of the noncommutative spacetime
\cite{lslett,lslong,lsem}.

\newsection{Automorphisms of Vertex Operator Algebras}

An automorphism of the vertex operator algebra $\alg$ is a unitary isomorphism
$\sigma:{\cal H}\to{\cal H}$ of vector spaces which preserves the
operator-state correspondence,
\beq
\sigma\,V(\psi;z_+,z_-)\,\sigma^{-1}=V(\sigma(\psi);z_+,z_-) \ .
\label{autodef}\eeq
In other words, the mapping $|\psi\rangle\to V(\psi;z_+,z_-)$ on ${\cal
H}\to\alg$ is equivariant with respect to the (adjoint) actions of $\sigma$ on
${\cal H}$ and $\alg$. The group of automorphisms \eqn{autodef} is denoted
${\rm Aut}(\alg)$. The grading of $\cal H$ with respect to conformal dimension
yields a decomposition of $\alg=\alg^+\otimes_{\complexs\{\Lambda\}}\alg^-$
into the eigenspaces of the Virasoro operators $L_0^\pm$ according to
$\alg^\pm=\mbox{$\bigoplus_{\Delta^\pm}$}\,\alg^\pm_{\Delta^\pm}$. The weight
zero subspace of $\cal H$ is one-dimensional and is spanned by the vacuum
$|{\rm
vac}\rangle$. The subgroup of ${\rm Aut}(\alg)$ consisting of automorphisms
that preserve this grading, i.e. $\sigma\alg_\Delta\sigma^{-1}=\alg_\Delta$, is
denoted ${\rm Aut}^{(0)}(\alg)$. The normal subgroup of ${\rm Aut}(\alg)$
consisting of inner automorphisms is denoted ${\rm Inn}(\alg)$, while the
remaining outer automorphisms are ${\rm Out}(\alg)={\rm Aut}(\alg)/{\rm
Inn}(\alg)$. The automorphism group is then the semi-direct product
\beq
{\rm Aut}(\alg)={\rm Inn}(\alg)~\semiprod{\rm Out}(\alg)
\label{fullautogp}\eeq
of ${\rm Inn}(\alg)$ by the natural action of ${\rm Out}(\alg)$.

The classification of the full automorphism group \eqn{fullautogp} is a
difficult problem in the general case. However, the structure of ${\rm
Aut}^{(0)}(\alg)$ can be determined to some extent. This group of automorphisms
is represented as a group of unitary transformations on each of the homogeneous
spaces ${\cal H}_\Delta$. Consider the chiral algebras $\alg^\pm$ built on the
compactification lattice $\Gamma$ with vertex operators $V(\psi_\pm;z_\pm)$,
where $\psi_\pm\in({\cal H}^\pm)^*$. Let $\psi_n$ be the component operators of
the meromorphic function $V(\psi;z_\pm)=\sum_{n\in\zeds}\psi_nz_\pm^{-n-1}$.
The condition \eqn{autodef} on $\alg^\pm$ is then equivalent to
$\sigma\psi_n\sigma^{-1}=(\sigma(\psi))_n~~\forall n\in\zed$. The automorphisms
of $\alg^\pm$ which preserve their gradings are those which leave invariant the
conformal vectors $\omega^\pm$ defined by the stress-energy tensors
$T^\pm(z_\pm)\equiv\sum_{k\in\zeds}L_k^\pm z^{-k-2}_\pm=V(\omega^\pm;z_\pm)$ of
the underlying conformal field theory. The operators $\omega^\pm$ are given
explicitly by
\beq
\omega^\pm=\mbox{$\frac12$}\,\id\otimes g_{\mu\nu}(\vec e^{\,\mu})_\lambda(\vec
e^{\,\nu})_\rho\alpha_{-1}^{(\pm)\lambda}\alpha_{-1}^{(\pm)\rho}\ ,
\label{confvecs}\eeq
where $\{\vec e^{\,\mu}\}$ is an arbitrary basis of the lattice $\Gamma$, and
they have conformal weight 2. The operations
\beq
[[\psi,\varphi]]\equiv\psi_0\varphi~~~~~~,~~~~~~
\langle\!\langle\psi,\varphi\rangle\!\rangle\equiv\psi_1\varphi
\label{Liedef1}\eeq
define, respectively, a Lie bracket and an invariant bilinear form on the space
\beq
{\cal L}_\Gamma^\pm=({\cal H}_1^\pm)^*/({\cal H}_1^\pm)^*\cap L_{-1}^\pm({\cal
H}^\pm_0)^*
\label{Liealg1}\eeq
of primary states of weight one. Then the unitary group $\exp i{\cal
L}_\Gamma^\pm$ yields a Lie group of automorphisms acting on the chiral algebra
$\alg^\pm$ by the adjoint representation\footnote{The $*$-conjugation on the
vertex operator algebra is defined by
$V^*(\psi;z_\pm)=\sum_{n\in\zeds}\psi^*_nz_\pm^{-n-1}=V(\e^{z_\pm
L_1^\pm}(-z_\pm^2)^{L_0^\pm}\psi;z_\pm^{-1})$. With this definition we have, in
particular, that $(L_n^\pm)^*=L_{-n}^\pm$.} and preserving its conformal
grading. These define the continuous automorphisms of the vertex operator
algebra, so that the inner automorphism subgroup of ${\rm Aut}^{(0)}(\alg^\pm)$
is
\beq
{\rm Inn}^{(0)}(\alg^\pm)={\rm Ad}_{\alg^\pm}\exp i{\cal L}_\Gamma^\pm \ .
\label{Inn0Apm}\eeq

Explicitly, ${\cal L}_\Gamma^\pm$ is generated by the operators
$\varepsilon_{q^\pm}\equiv(-1)^{q_\mu w^\mu}\e^{-iq_\mu^\pm x_\pm^\mu}$ (the
generators of the twisted group algebra $\complex\{\Gamma\}$ of $\Gamma$),
where $q^\pm\in\Gamma_2\equiv\{q\in\Gamma~|~q_\mu g^{\mu\nu}q_\nu=2\}$, and
$r^\pm_\mu\alpha_{-1}^{(\pm)\mu}$, where $r^\pm\in\Gamma$. As a subspace of the
noncommutative spacetime, the algebra \eqn{Liealg1} therefore contains the
lowest-lying non-trivial oscillatory modes of the strings (i.e. the graviton
operators \eqn{graviton}), and thus the smallest quantum perturbation of
classical spacetime. With the Lie bracket in \eqn{Liedef1} its commutation
relations are
\beq\new{\begin{array}{c}
\left[\left[\alpha_{-1}^{(\pm)\mu},\alpha_{-1}^{(\pm)\nu}\right]\right]
=0~~~~~~,~~~~~~\left[\left[\alpha_{-1}^{(\pm)\mu},\varepsilon_{q^\pm}
\right]\right]=g^{\mu\nu}q_\nu^\pm~
\varepsilon_{q^\pm}\\\left[\left[\varepsilon_{q^\pm},\varepsilon_{r^\pm}
\right]\right]=\left\{\new{\begin{array}{ll}0&,~~q_\mu^\pm
g^{\mu\nu}r_\nu^\pm\geq0\\(-1)^{r_\mu
v^\mu}~\varepsilon_{q^\pm+r^\pm}&,~~q_\mu^\pm
g^{\mu\nu}r_\nu^\pm=-1\\q_\mu^\pm\alpha_{-1}^{(\pm)\mu}&,~~q_\mu^\pm
g^{\mu\nu}r_\nu^\pm=-2\end{array}}\right.\end{array}}
\label{LGamma}\eeq
Using the bilinear form in \eqn{Liedef1} we have
$\langle\!\langle\alpha_{-1}^{(\pm)\mu},\alpha_{-1}^{(\pm)\nu}\rangle\!\rangle
=g^{\mu\nu}$,
$\langle\!\langle\alpha_{-1}^{(\pm)\mu},\varepsilon_{q^\pm}\rangle\!\rangle=0$,
and $\langle\!\langle\varepsilon_{q^\pm},\varepsilon_{r^\pm}\rangle\!\rangle=0$
if $q_\mu^\pm g^{\mu\nu}r_\nu^\pm\geq-1$, 1 if $q_\mu^\pm
g^{\mu\nu}r_\nu^\pm=-2$. This determines a root space decomposition of ${\cal
L}_\Gamma^\pm$ such that its root lattice is precisely $\Gamma$ and its set of
roots is $\Gamma_2$. Furthermore, we can use this bilinear form and the
operator-state correspondence to construct a level 1 representation of the
corresponding affinization $\widehat{\cal L}^\pm_\Gamma$, appropriate to the
action of \eqn{Inn0Apm} on the quantum fields of the string theory. One finds
for the usual modes that $(\alpha_{-1}^{(\pm)\mu})_n=\alpha_n^{(\pm)\mu}$ and
$(\varepsilon_{q^\pm})_n$ can be determined from the tachyon vertex operators
\cite{lslong}, so that
\beq\new{\begin{array}{c}
\left[\left[(\alpha_{-1}^{(\pm)\mu})_m,(\alpha_{-1}^{(\pm)\nu})_n\right]
\right]=mg^{\mu\nu}\delta_{m+n,0}~~~~~~,~~~~~~
\left[\left[(\alpha_{-1}^{(\pm)\mu})_m,(\varepsilon_{q^\pm})_n\right]
\right]=g^{\mu\nu}q_\nu^\pm~
(\varepsilon_{q^\pm})_{m+n}\\\left[\left[(\varepsilon_{q^\pm})_m,
(\varepsilon_{r^\pm})_n\right]\right]
=\left\{\new{\begin{array}{ll}0&,~~q_\mu^\pm
g^{\mu\nu}r_\nu^\pm\geq0\\(-1)^{r_\mu
v^\mu}~(\varepsilon_{q^\pm+r^\pm})_{m+n}&,~~q_\mu^\pm
g^{\mu\nu}r_\nu^\pm=-1\\q_\mu^\pm(\alpha_{-1}^{(\pm)\mu})_{m+n}
+m\delta_{m+n,0}&,~~q_\mu^\pm
g^{\mu\nu}r_\nu^\pm=-2\end{array}}\right.\end{array}}
\label{LGammaaff}\eeq
With the Lie bracket defined in \eqn{Liedef1}, the action of the affine Lie
algebra $\widehat{\cal L}_\Gamma^\pm$ on $\alg^\pm$ can be represented by the
actions of the corresponding vertex operators in the ordinary commutator
bracket.

The Kac-Moody algebras $\widehat{\cal L}_\Gamma^\pm$ always contain the affine
$u(1)_\pm^d$ gauge groups generated by the chiral Heisenberg fields
\beq
\alpha_\pm^\mu(z_\pm)=-i\partial_{z_\pm}X_\pm^\mu(z_\pm)=
\sum_{k=-\infty}^\infty\alpha_k^{(\pm)\mu}\,z_\pm^{-k-1}=-i
V(\id\otimes\alpha_{-1}^{(\pm)\mu};z_\pm)
\label{u1currents}\eeq
which are the conserved currents associated with the target space
reparametrization symmetry. The full algebraic structure of ${\cal
L}_\Gamma^\pm$ is, however, strongly dependent on the particular
compactification lattice $\Gamma$. For instance, when $\Gamma=(2\zed)^d$, for
which $\Gamma_2=\emptyset$, we have ${\cal L}_{(2\zeds)^d}^\pm=u(1)_\pm^d$,
while for $\Gamma=\zed^d$ (corresponding to a complete factorization of the
$d$-torus as $T^d=(S^1)^d$), we find ${\cal L}_{\zeds^d}^\pm=su(2)_\pm^d$.

{}From \eqn{confvecs} it is also possible to immediately identify the outer
automorphism subgroup of ${\rm Aut}^{(0)}(\alg_\pm)$ as the (discrete) isometry
group of the lattice $\Gamma$,
\beq
{\rm Out}^{(0)}(\alg_\pm)={\rm Aut}(\Gamma)=O(d;\zed) \ ,
\label{out0Apm}\eeq
which represents the symmetries of the Dynkin diagram of ${\cal L}_\Gamma^\pm$.
However, a complete classification of the full automorphism group, including
the transformations which do not preserve \eqn{confvecs}, is not as
straightforward. For instance, the inner automorphism of $\alg^\pm$ determined
by the unitary element
\beq
\sigma_{q^\pm}=\exp\left\{2i\,\varepsilon_{q^\pm}\otimes\id/(2-q_\mu^\pm
g^{\mu\nu}q_\nu^\pm)\right\}
\label{sigmaqdef}\eeq
for $q^\pm\in\Gamma-\Gamma_2$ maps \eqn{confvecs} to the conformal
vector $\omega_{q^\pm}^\pm=\sigma_{q^\pm}\omega^\pm\sigma_{q^\pm}^{-1}
=\omega^\pm+L_{-1}^\pm(\varepsilon_{q^\pm}\otimes\id)$ of conformal dimension
$\frac12q_\mu^\pm g^{\mu\nu}q_\nu^\pm+1$. Thus the full group of inner
automorphisms in this case is quite large. The classification of the
automorphism group for the full vertex operator algebra
$\alg=\alg^+\otimes_{\complexs\{\Lambda\}}\alg^-$ is further complicated by the
existence of transformations which do not factorize into chiral components and
can mix between the two chiral algebras.

\newsection{Quantum Geometry and Gauge Symmetries of String Theory}

We shall now apply the formalism developed thus far to a systematic
investigation of the symmetries of the noncommutative string spacetime,
represented by the $*$-algebra
$\alg=\alg^+\otimes_{\complexs\{\Lambda\}}\alg^-$. As discussed in section 1,
the inner automorphisms of an algebra in Noncommutative Geometry
represent gauge tranformations. In simple cases (such as the standard model),
the distinction between outer and inner automorphisms is, respectively, of
diffeomorphisms of the spacetime and internal gauge symmetries. In the case at
hand, however, the situation is far from being so simple, as the entire algebra
is noncommutative and a manifold only emerges in a low energy limit.

The quantum geometry of $T^d$ is determined by classifying its duality
symmetries. The basic observation is that the entire algebraic structure of the
vertex operator algebra is unchanged on the whole by redefining the Heisenberg
fields \eqn{u1currents} as $\alpha_\pm^\mu(z_\pm)\to\pm\alpha_\pm^\mu(z_\pm)$.
This transformation can be achieved via an automorphism of $\alg$ in several
ways. Such automorphisms define the duality transformations of the string
theory and represent symmetries of the classical spacetime $T^d$ when viewed as
a (low-energy) subspace of the full noncommutative spacetime. The simplest
example is the T-duality transformation $d^\pm\to(d^\pm)^{-1}$, which
corresponds essentially to an inversion of the spacetime metric $g_{\mu\nu}$
and interchanges momentum and winding modes in the spectrum of the quantum
string theory. It implies that, as subspaces of the noncommutative spacetime,
the torus $T^d$ is equivalent to its dual $(T^d)^*=\real^d/2\pi\Gamma^*$. From
the point of view of classical general relativity there is no reason for these
two spacetimes to describe the same physics, and the noncommutative geometry
naturally describes the stringy modification of classical spacetime geometry.

The augmentation of the generic $u(1)_+^d\oplus u(1)_-^d$ gauge symmetry
of the string theory to ${\cal L}_\Lambda\equiv{\cal L}_\Gamma^+\oplus{\cal
L}_\Gamma^-$ is known as an `enhanced gauge symmetry'. For the full vertex
operator algebra $\alg$, it is due to the appearence of extra dimension
$(\Delta^+,\Delta^-)=(1,0)$ and $(0,1)$ operators in the theory (namely the
appropriate tachyon fields). Such operators can be used to perturb the
underlying conformal field theory to an isomorphic one with different target
space properties. It turns out that to describe the full duality group of
toroidally compactified string theory, it suffices to examine the Lie algebra
associated with the unique fixed point $(d^\pm)^2=\id$ of the T-duality
transformation, i.e. $g_{\mu\nu}=\delta_{\mu\nu},\beta_{\mu\nu}=0$, or
equivalently $\Gamma=\zed^d$. As mentioned in the previous section, at this
fixed point the generic $U(1)_+^d\times U(1)_-^d$ gauge symmetry is enhanced to
a level 1 representation of the affine Lie group
$\widehat{SU(2)}_+^d\times\widehat{SU(2)}_-^d$. To describe this structure
explicitly, let $k_\mu^{(i)}$, $i=1,\dots,d$, be a suitable basis of (constant)
Killing forms on $T^d$. Then the chiral vertex operators
$J_\pm^{\eta(i)}(z_\eta)=:\e^{\pm ik_\mu^{(i)}X_\eta^\mu(z_\eta)}:,
J_3^{\eta(i)}(z_\eta)=ik_\mu^{(i)}\alpha_\eta^\mu(z_\eta)$, with $\eta=\pm$,
generate a level 1 $su(2)_+^d\oplus su(2)_-^d$ Kac-Moody algebra as described
before. Associated with this gauge symmetry of the theory are the corresponding
gauge group elements $\sigma=\e^{i{\cal G}}$. The generators $\cal G$ that
implement spacetime duality transformations of the string theory were
originally constructed in \cite{strsymsdual} and are given as follows,
\begin{eqnarray}
{\cal G}^{(\mu)}(z_+,z_-)&=&\mbox{$\frac\pi{2i}$}~:~\e^{i\sqrt2\,k_\mu^{(\mu)}
X^\mu(z_+,z_-)}-\e^{-i\sqrt2\,k_\mu^{(\mu)}X^\mu(z_+,z_-)}~:\label{genmirror}
\\{\cal
G}_\pm^{(\mu)}(z_+,z_-)&=&\mbox{$\frac\pi{2i}$}~:~\e^{iX_\pm^\mu(z_\pm)}
-\e^{-iX_\pm^\mu(z_\pm)}~:\label{genrefl}\\
{\cal
G}_\lambda(z_+,z_-)&=&\lambda_\mu(X)\left(z_+\alpha_+^\mu(z_+)
-z_-\alpha_-^\mu(z_-)\right)\label{genbeta}\\{\cal
G}_\xi(z_+,z_-)&=&\xi_\mu(X)\left(z_+\alpha_+^\mu(z_+)
+z_-\alpha_-^\mu(z_-)\right)
\label{gendiff}\end{eqnarray}
where $X=\frac1{\sqrt2}(X_++X_-)$.

The operator \eqn{genmirror} generates the $\mu^{\rm th}$ factorized duality
map of the spacetime. For each $\mu=1,\dots,d$ it is a generalization of the
$R\to1/R$ circle duality in the $X^\mu$ direction (of radius $R$) of $T^d$.
When $d$ is even, it corresponds to mirror symmetry which interchanges the
complex and K\"ahler structures of the target space and leads to the stringy
phenomenon of spacetime topology change. The inner automorphism generated by
${\cal G}_+^{(\mu)}+{\cal G}_-^{(\mu)}$ corresponds to the reflection
$X^\mu\to-X^\mu$ of the coordinates of $T^d$. Thus factorized dualities and
spacetime reflections are enhanced gauge symmetries, and hence intrinsic
properties, of the noncommutative spacetime. The remaining two duality
transformations are abelian gauge symmetries. The operator \eqn{genbeta}
generates local gauge transformations $\beta\to\beta+d\lambda$ of the instanton
two-form. Taking $\lambda_\mu(X)=C_{\mu\nu}X^\nu$, with $C_{\mu\nu}$ a constant
antisymmetric matrix, gives the shift
$\beta_{\mu\nu}\to\beta_{\mu\nu}+C_{\mu\nu}$. Singlevaluedness of the
corresponding gauge group element $\sigma$ requires $C_{\mu\nu}\in\zed$
yielding a large gauge transformation. Finally, the operator \eqn{gendiff}
generates a general spacetime coordinate transformation $X\to\xi(X)$. Again for
the large diffeomorphisms $\xi_\mu(X)=T_{\mu\nu}X^\nu$ of $T^d$,
singlevaluedness requires $[T_{\mu\nu}]\in SL(d,\zed)$. In particular, setting
$T_{\mu\nu}=\frac\pi2{\rm
sgn}(P)g_{P(\mu),\nu}$ yields a permutation $P\in S_d$ of the coordinates
of $T^d$. Combining these with the factorized duality transformations yields
T-duality in the form of an inner automorphism. As such, T-duality corresponds
to the global gauge transformation in the Weyl subgroup $\zed_2$ of $SU(2)$. It
can be shown that these discrete vertex operator inner automorphisms,
corresponding to large gauge transformations of the internal string spacetime,
generate the infinite discrete duality group $O(d,d;\zed)$ which is the
isometry group of the Narain lattice $\Lambda$. Explicit expressions for the
actions of these operators on $\cal H$ and $\alg$ can be found in
\cite{lslong}. We see therefore that, by viewing string theory as a
noncommutative geometry, target space duality transformations, as well as
generic diffeomorphisms, all appear naturally as elements of the group of gauge
transformations.

It is also possible to write down the full outer automorphism group,
representing the diffeomorphisms (i.e. the gravitational symmetries) of the
noncommutative spacetime, as
\beq
{\rm Out}(\alg)=O(d,d;\zed)~\semiprod O(2) \ .
\label{outerfull}\eeq
The $O(2)$ part of \eqn{outerfull} is a worldsheet symmetry group that acts by
rotating the two chiral sectors into each other. It arises from the spin
structure of the string worldsheet which yields a representation of
$spin(2)$ on the Hilbert space $\cal H$ that implements the group
$SO(2)$. In particular, its $\zed_2$ subgroup generates the
worldsheet parity transformation that interchanges the left and right chiral
algebras, $\alg^+\otimes\alg^-\to\alg^-\otimes\alg^+$. However, as discussed
earlier, it is not known what the general form of the inner automorphism group
is. It is an open problem to
determine what symmetries are represented by inner automorphisms such as
\eqn{sigmaqdef} which correspond to deformations of the underlying conformal
field theory to an inequivalent two-dimensional quantum field theory.

All of these duality transformations are exact symmetries of the full
noncommutative spacetime described by the entire vertex operator algebra. The
algebra $C^\infty(T^d)$
which describes the toroidal spacetime is contained in the (smeared) vertex
operator algebra as a subalgebra. The generators \eqn{genmirror}--\eqn{genbeta}
act trivially on the low-energy sector, i.e. $\e^{i{\cal
G}}|_{C^\infty(T^d)}=\id$, as expected since the symmetries represented by the
inner automorphisms \eqn{Inn0Apm} live in the quantum perturbation of classical
spacetime represented by the graviton operators \eqn{graviton}. The main
feature is that the orthogonal projection $\alg\to C^\infty(T^d)$ does not
commute with the duality maps, and therefore these symmetries change the
geometry and topology of the classical spacetime. Thus although duality is a
gauge symmetry of the full noncommutative geometry, the low energy sectors look
dramatically different. Similar statements also apply to the outer
automorphisms \eqn{outerfull} acting on $C^\infty(T^d)$.

As for the inner automorphisms generated by \eqn{gendiff}, when projected onto
the subalgebra $C^\infty(T^d)\subset\alg$ they represent the generators of
${\rm Diff}(T^d)$ in terms of the canonically conjugate center of mass
variables $x^\mu,p_\mu$. Thus the full group of {\it inner} automorphisms of
$\alg$, representing internal gauge symmetries of the string spacetime,
projects onto the group of {\it outer} automorphisms of $C^\infty(T^d)$,
representing the diffeomorphisms of the target space. In this respect general
covariance is represented as a gauge symmetry in the stringy modification of
general relativity, in that its transformations are generated by a unitary
group acting on a noncommutative geometry. This gives a remarkable
interpretation of the usual classical symmetries in quantum geometry, a
dramatic example of which is that gravity becomes a gauge theory.

\newsection{Universal Gauge Symmetry}

We have seen that Noncommutative Geometry gives a unifying description of
spacetimes that appear to be distinct at low energies, as well as of gauge
transformations and diffeomorphisms. It is natural then to envisage some sort
of {\em universal} gauge symmetry which encompasses all of these features. The
investigation of such a universal symmetry has been previously attempted in
\cite{rajeev1}. There the Universal Gauge Group is defined as an ideal in the
algebra of bounded operators acting on a Hilbert space\footnote{An extension to
large-$N$ matrix models for open strings has also been recently given in
\cite{rajeev2} involving some more familiar algebras such as
$gl(\infty,\complex)$ and the Cuntz algebra.}, which contains {\it all} groups
of gauge tranformations as subgroups. These ideals are related to the Schatten
ideals which play a central role in the description of infinitesimals in
noncommutative geometry \cite{Book}, and this construction can be carried out
using the notion of a cycle in noncommutative geometry. This group is easier to
deal with than the inductive large-$N$ limit $U(\infty)$ of the unitary groups
$U(N)$, which also contains all compact gauge groups as subgroups (via
appropriate unitary representations). Two Yang-Mills theories with different
structure groups can have vastly different physical properties, so that this
notion of a universal gauge theory overlies many different physical theories.
In the present context we can make a construction similar in spirit to this.

We recall the dependence of the symmetry group \eqn{Inn0Apm} on the
choice of compactification lattice $\Gamma$. Since nature cannot distinguish
between different compactifications (as this part of the full string spacetime
is unobservable in the physical spacetime in which we live), we seek a symmetry
algebra which overlies {\it every} Lie algebra ${\cal L}_\Lambda$.
This would ultimately lead to a unified framework for studying all of the
dynamical symmetries of string theory. For this, we consider the unique
$2d$-dimensional even self-dual Lorentzian lattice ${\mit\Pi}_{d,d}$, and we
analytically continue in the spacetime momenta to allow for both purely real
and purely imaginary momenta by extending it to a module over the Gaussian
integers, $\Lambda^{\rm (G)}\equiv{\mit\Pi}_{d,d}\otimes_\zeds\zed[i]$. This
extension will ensure that the roots of the appropriate Lie algebra lie inside
the light-cone of the Lorentzian lattice $\Lambda^{\rm (G)}$. Now we carry out
the constructions of sections 2 and 3 with the arbitrary lattice
$\Lambda=\Gamma^*\oplus\Gamma$ replaced by $\Lambda^{\rm (G)}$. Then the
corresponding Lie algebra of dimension 1 primary states
\beq
{\cal L}_U\equiv({\cal H}_1^{\rm (G)})^*/\,\mbox{$\bigcup_{k\geq1}$}\,({\cal
H}_1^{\rm (G)})^*\cap\left(L_{-k}^+\otimes L_{-k}^-\right)({\cal H}^{\rm
(G)})^*
\label{universalLiealg}\eeq
generates a {\it universal} gauge symmetry group, in the sense that for any
compactification lattice $\Gamma$ there is a natural Lie subalgebra embedding
${\cal L}_\Lambda\hookrightarrow{\cal L}_U$ \cite{moore1}. In this way,
noncommutative geometry yields a natural geometrical interpretation to the
universal gauge group corresponding to \eqn{universalLiealg}. This
construction, as well as those of \cite{rajeev1,rajeev2}, is based on the
representation of gauge tranformations as operators on an infinite-dimensional
Hilbert space.

The universal gauge symmetry algebra \eqn{universalLiealg} (as well as that of
\cite{rajeev2}) is a generalization of the Lie algebra of vector fields on the
circle to noncommutative geometry. These structures are all related to the
symmetries of the noncommutative 2-torus with rational deformation parameter
$\theta$, which in turn is the underlying noncommutative geometrical structure
of M Theory. The Lie group $U(N)$ is just the unitary group of the
$C^*$-algebra
$M(N,\complex)$. Thus, given the relation between the vertex operator algebra
and the large-$N$ limit of this noncommutative torus, the symmetry group ${\rm
Aut}(\alg)$ and the corresponding universal gauge group should be related in
some way to those of the infinite-dimensional $C^*$-algebra
$M(\infty,\complex)$. Thus the universal gauge symmetries described here could
help in understanding the underlying dynamical symmetries of M Theory.

Another feature which arises in this context concerns the meaning of the
gauge symmetries in ${\rm Inn}(\alg)-{\rm Inn}^{(0)}(\alg)$. Such
transformations represent higher-order quantum perturbations of classical
spacetime and can dramatically alter the underlying world-sheet theory. The
symmetry group ${\rm Aut}(\alg)$ then overlies a rather large set of
two-dimensional quantum field theories which can be physically quite different.
But from the point of view of noncommutative geometry, these theories are all
embedded into the same universal structure and are just different corners of
some big model whose symmetry group is ${\rm Aut}(\alg)$. The different
corners are related to one another by gauge transformations, yielding duality
maps between inequivalent physical theories. This is the earmark of M Theory,
which overlies the five consistent superstring theories in ten dimensions by
relating them to each other via duality isomorphisms. In this sense, the
noncommutative geometry of string spacetimes is naturally suited to the
unifying framework of M Theory. The focal point of the ease in which these
interpretations arise is at the very heart of the techniques of noncommutative
geometry.

\newsection{Time-like Compactifications and Generalized Kac-Moody Symmetries}

The structures we have described thus far change rather dramatically when the
toroidal target space contains a time-like coordinate, i.e. $T^{1,d-1}\cong
S_-^1\times T^{d-1}$, generated by a compactification lattice $\Gamma$ of
Minkowski signature. It was shown in \cite{moore1} that the action of the
duality group ${\rm Aut}(\Lambda)$ on the background parameters of the
spacetime is ergodic and dense in this case. This means that the quantum moduli
space of time-like toroidal compactifications does not exist as a manifold and
only makes sense within the framework of noncommutative geometry. This was
precisely the case of the noncommutative torus with irrational deformation
parameter $\theta$, the duality action being regarded there as the orbit of a
free particle in the 2-torus. Noncommutative geometry is therefore an important
ingredient in the description of the space of time-dependent string
backgrounds.

The structure of the vertex operator algebra also changes when the lattice
$\Gamma$ is no longer Euclidean. For example, consider the unique
26-dimensional even self-dual Lorentzian lattice $\Gamma={\mit\Pi}_{1,25}$.
Then $\Lambda_*={\mit\Pi}_{1,25}\oplus{\mit\Pi}_{1,25}$ is the unique point in
the quantum moduli space of time-like toroidal compactifications at which the
vertex operator algebra $\alg$ completely factorizes between its left and right
chiral sectors,
\beq
{\cal H}(\Lambda_*)^*={\cal C}^+\otimes{\cal C}^-
\label{H*fact}\eeq
where
\beq
{\cal C}^\pm=\complex\{{\mit\Pi}_{1,25}\}\otimes{\cal
F}^\pm({\mit\Pi}_{1,25})^*
\label{Cpm}\eeq
and we have denoted the explicit dependence of the Heisenberg algebras built on
${\mit\Pi}_{1,25}$. In essence this factorizes the closed string into two open
strings. The distinguished lattice $\Lambda_*$ is an enhanced symmetry point in
the moduli space. The corresponding Lie algebra ${\cal L}_{\Lambda_*}={\cal
B}\oplus{\cal B}$ is a {\it maximal} symmetry algebra, in the sense that it
contains all gauge symmetry algebras \cite{moore1}. It is not, however,
universal because the gauge symmetries are not necessarily embedded into it as
Lie subalgebras. ${\cal L}_{\Lambda_*}$ contains many novel
infinite-dimensional symmetry algebras, such as algebras of area-preserving and
volume-preserving diffeomorphisms. The existence of these large symmetries is
intimately connected to the unusual nature of time in string theory.

The chiral algebra $\cal B$ obtained from the complete factorization above is
given explicitly by
\beq
{\cal B}={\cal
H}_1^\pm({\mit\Pi}_{1,25})^*/\ker\langle\!\langle\cdot,\cdot\rangle\!
\rangle
\label{fakemonster}\eeq
where in the present case we have to divide out by the kernel of the bilinear
form in \eqn{Liedef1} because in Minkowskian signature additional null physical
states besides the spurious states of $({\cal H}_1^\pm)^*\cap L_{-1}^\pm({\cal
H}^\pm_0)^*$ appear. $\cal B$ is called the fake Monster Lie
algebra \cite{flm,borcherds}. It is {\it not} a Kac-Moody algebra because it
contains light-like simple Weyl roots. This leads to the notion of a Borcherds
or generalized Kac-Moody algebra \cite{borcherds} which resembles an ordinary
affine Lie algebra in most respects, except for the existence of imaginary
(non-positive norm) simple roots. The positive norm simple roots of the root
lattice ${\mit\Pi}_{1,25}$ of $\cal B$ lie in the Leech lattice $\Gamma_{\rm
L}$, the unique 24-dimensional even self-dual Euclidean lattice with no
vectors of length $\sqrt2$.

The fake Monster Lie algebra is the starting point for the construction of
another generalized Kac-Moody algebra, the Monster Lie algebra \cite{flm}. The
Monster module $\cal M$ is built from the Leech lattice
$\Gamma_{\rm L}$ and the order 2 automorphism $\sigma$ of the corresponding
lattice vertex operator algebra induced by the reflection isometry of
$\Gamma_{\rm L}$. $\cal M$ is the vertex operator algebra associated with the
(unique) chiral $c=24$ $\sigma$-twisted orbifold conformal field theory
\cite{flm}. Since ${\mit\Pi}_{1,25}\cong\Gamma_{\rm
L}\oplus{\mit\Pi}_{1,1}$, with ${\mit\Pi}_{1,1}$ the unique two-dimensional
even self-dual Lorentzian lattice, $\cal M$ is naturally obtained from $\cal
B$. The most startling aspect of $\cal M$ is that it contains {\it no}
dimension 1 operators. Its symmetries (which are all discrete) are now
classified according to the space of dimension 2 primary states. On this space
there is a binary operation
\beq
\psi\star\varphi\equiv\psi_1\varphi
\label{starprod}\eeq
which turns it into a commutative non-associative algebra. This algebra is the
196884-dimensional Griess algebra for which the Monster group $F_1$ is the full
automorphism group. $F_1$ is the largest finitely-generated simple sporadic
group.

The monster module $\cal M$ is naturally associated with a twisted heterotic
string theory \cite{harvey}. It leads to a rather exotic noncommutative
spacetime which possesses only discrete outer automorphisms as its
(diffeomorphism) symmetries,  a subgroup of which is the Monster group. This
spacetime is ``topological" in the sense that it does not contain the usual
(low-energy) diffeomorphism symmetries. It can also be shown that string
compactifications with Monster symmetry are not dense \cite{moore1}.

It has been recently argued \cite{chapline} that various superstring dualities
are isomorphic to the elementary group structures occuring above in the
construction of $F_1$. In particular, string-string duality appears to be
identical to the reflection isometry of the Leech lattice, and an explicit
$SU(\infty)$ gauge symmetry (the symmetry group of the 11-dimensional
supermembrane) arises from this formulation whose fundamental quantum degrees
of freedom can be identified with the elements of the Griess algebra. The
emergence of matter-like states is thus an example of enhanced gauge symmetry.
Thus the Monster sporadic group appears to be a hidden symmetry group of
superstring theory\footnote{In this paper we have discussed only the simplest
case of bosonic strings, but, as is discussed in
\cite{fg}--\cite{lslong},\cite{lsem}, the natural arena for the noncommutative
geometry of quantum field theories is appropriate supersymmetric extensions of
these models.}. The Monster module has also been related recently to D-particle
dynamics in type IIA superstring theory \cite{gk}. In \cite{moore2} the
relations between string duality and generalized Kac-Moody algebras are also
discussed. The space of perturbative BPS string states of toroidally
compactified heterotic string theories forms a generalized Kac-Moody algebra.
The possibility of relating superstring theory to properties of exotic
algebras, such as the Leech lattice and the Monster group, is quite appealing
since many of the associated symmetry groups are unique.

Given these relations and the fact that the fake Monster Lie algebra
\eqn{fakemonster} is a maximal symmetry algebra of the string theory, it is
becoming increasingly evident that Borcherds algebras, when interpreted as
generalized symmetry algebras of the noncommutative geometry, are relevant to
the construction of a universal symmetry of string theory. Being natural
generalizations of affine Lie algebras, they may emerge as new symmetry
algebras for string spacetimes. The appearence of exotic gauge symmetries such
as Monster symmetry could have significant implications as the role of a
dynamical Lie algebra representing the unified symmetries of the noncommutative
string spacetime. The remarkable fact is that the irreducible representations
of the algebras described above match exactly the Fock space of the bosonic
string
field theory with the underlying Kac-Moody algebra as spectrum-generating
algebra for the one-string Hilbert space. Within these dynamical algebras then
one would expect to find all quantum states in a single representation, with
the underlying Lie algebra determining the maximal symmetry algebra of the
string spacetime.

\bigskip

\noindent
{\bf Acknowledgements:} We thank G. Mason for correspondence concerning vertex
operator algebras. The work of {\sc r.j.s.} was supported in part by PPARC
(U.K.).

\end{document}